\documentclass[showpacs,preprintnumbers,amsmath,amssymb]{revtex4}

\usepackage{graphicx}
\usepackage{dcolumn}
\usepackage{bm}
\begin{document}
\title{Neutrino nucleus cross sections}
\author{M. Sajjad Athar, S. Chauhan, S. K. Singh}
\affiliation{Department of Physics, Aligarh Muslim University, \\Aligarh-202 002, India.}
\author{M. J. Vicente Vacas}
\affiliation{Departamento de F\'{\i}sica Te\'orica and IFIC, \\
Centro Mixto Universidad de
Valencia-CSIC,\\
46100 Burjassot (Valencia), Spain}
\date{\today}
\begin{abstract}
We present the results of our calculation which has been performed to study the nuclear effects in the quasielastic, inelastic and deep inelastic scattering of neutrinos(antineutrinos) from nuclear targets. These calculations are done in the local density approximation. We take into account the effect of Pauli blocking, Fermi motion, Coulomb effect, renormalization of weak transition strengths in the nuclear medium in the case of the quasielastic reaction. The inelastic reaction leading to production of pions is calculated in a $\Delta $- dominance model taking into account the renormalization of $\Delta$ properties in the nuclear medium and the final state interaction effects of the outgoing pions with the residual nucleus. We discuss the nuclear effects in the $F_{3}^{A}(x)$ structure function in the deep inelastic neutrino(antineutrino) reaction using a relativistic framework to describe the nucleon spectral function in the nucleus.
\end{abstract}
\maketitle
It is now well established that neutrinos oscillate. The next target of the experimentalist is to determine precisely the various parameters of neutrino mass matrix given by Pontecarvo-Maki-Nakagawa-Sakata (PMNS), absolute masses of different flavors of neutrinos, pattern of the known neutrino mass differences, CP violation in neutrino sector, etc. To address some of these problems several experiments like CNGS, MINOS and SciBooNE are taking data. The eperimental analyses of neutrino oscillation data are going on, for example, at K2K and MiniBooNE, and several experiments are planned to be done in future like T2K and NO$\nu$A. Besides the accelerator experiments, experiments with neutrinos from $\nu$- factories, $\beta$-beams, etc., are also planned, as well as some experiments with natural $\nu$-sources like solar neutrinos, atmospheric neutrinos, or (anti)neutrinos from nuclear reactors are also planned. These experiments use various nuclear targets like $^{12}C$ by MiniBooNE, SciBooNE, MINER$\nu$A and NO$\nu$A, $^{16}O$ by SuperKamiokande, T2K, UNO, Hyper-K, K2K and MEMPHYS, $^{40}Ar$ by ICARUS and NO$\nu$A, $^{56}Fe$ by MINOS, INO, MINER$\nu$A, and $^{208}Pb$ by CNGS, MINER$\nu$A and OPERA collaborations. Most of these experiments are being done in the neutrino energy region of $E_{\nu}<2 GeV$, for example at MiniBooNE, the average energy($<E_{\nu}>$) is 750 MeV while at K2K it is 1.3 GeV. At the energies of a few GeV the contribution to the cross section comes from the quasielastic, inelastic and deep inelastic processes and the analysis of the data of neutrino experiments is based on Monte Carlo generator of events like NUANCE, NEUGEN, NEUT, etc. In these Monte Carlo generators, neutrino cross sections are used which are based on the model of Llewellyn Smith~\cite{Smith} and Smith and Moniz~\cite{Moniz} for the quasielastic reactions, Rein and Sehgal ~\cite{Rein} for the inelastic reactions and GRV98\cite{Grv98} along with the modifications suggested by Bodek et al.~\cite{Bodek} for the deep inelastic reactions. 

The importance of a better knowledge of neutrino-nucleus cross section to be used in the Monte Carlo generators has been realized and discussed in a series of neutrino conferences like NuInt, NuFact, NOW, etc. There are, now, various theoretical calculations for the quasielastic process which make use of nuclear models like shell model with pairing correlations, random phase approximation, relativistic mean field approximation, etc. In the case of inelastic neutrino-nuclear reactions pion production processes have been studied. These are generally studied in a $\Delta$-dominance model in which pions are dominantly produced through the excitation of $\Delta$ and its subsequent decay leading to pions. Some of the calculations have also been done by taking background as well as higher resonance terms. In the case of deep inelastic scattering of neutrinos(antineutrinos) from nuclear targets, there are very few calculations where the dynamical origin of the nuclear medium effects have been studied. In some phenomenological analyses, nuclear medium effects have been described in terms of few parameters which are determined from fitting the experimental data of charged leptons and neutrino(antineutrino) deep inelastic scattering from various nuclear targets.

Here we discuss the nuclear effects in the quasielastic reaction, inelastic one $\pi$ production in the $\Delta$ dominance model and the nuclear effects on the $F_{3}^{A}(x,Q^2)$ structure function in the deep inelastic reaction. These calculations are done in the local density approximation. For the quasielastic process this model takes into account the effect of Pauli blocking, Fermi motion, Coulomb effect, renormalization of weak transition strengths in the nuclear medium. The inelastic reaction leading to production of pions is calculated in a $\Delta $- dominance model taking into account the renormalization of $\Delta$ properties in the nuclear medium and the final state interaction effects on the outgoing pions. We discuss the nuclear effects in the $F_{3}^{A}(x)$ structure function in the deep inelastic neutrino(antineutrino) reaction using a relativistic framework to describe the nucleon spectral function in the nucleus. The details of these calculations may be found in the Refs.~\cite{Singh1,Athar2} for the quasielastic process, Refs.~\cite{Singh2,Athar3} for the inelastic process and Ref.~\cite{Athar5} for the deep inelastic process. Similar calculations for nuclear effects in the quasielastic and inelastic processes have also been recently done by many other groups~\cite{Nieves}-\cite{Amaro}. 
In the following we describe, in brief, the formalism for calculating the nuclear effects in quasielastic, inelastic and deep inelastic processes in Sec.1, and present the numerical results in Sec.2 with concluding remarks given in Sec.3.   
\section{NEUTRINO NUCLEUS REACTIONS}
\subsection{QUASIELASTIC REACTION}
The basic reaction for the quasielastic process is a neutrino interacting with a neutron inside the nucleus which is given by
\begin{equation}\label{quasi_reaction}
\nu_{\mu}(k) + n(p) \rightarrow \mu^{-}(k^{\prime}) + p(p^{\prime})
\end{equation}
The cross section for quasi-elastic charged lepton production is calculated in the local density approximation by taking into account the Fermi motion and the Pauli blocking effects through the imaginary part of the Lindhard function
for the particle hole excitations in the nuclear medium. The
renormalization of the weak transition strengths are calculated in the random
phase approximation(RPA) through the interaction of the p-h excitations as
they propagate in the nuclear medium using a nucleon-nucleon potential
described by pion and rho exchanges. The effect of the Coulomb distortion
of muon in the field of final nucleus is taken into account
using a local version of the modified effective momentum
approximation. 

The total cross section $\sigma(E_\nu)$ for the charged current neutrino induced reaction on a nucleon inside the nucleus in a local Fermi gas model is written as~\cite{Athar2}:
\begin{eqnarray}\label{sigma_quasi}
\sigma(E_\nu)&=&-\frac{2{G_F}^2\cos^2{\theta_c}}{\pi}\int^{r_{max}}_{r_{min}} r^2 dr \int^{p_\mu^{max}}_{p_\mu^{min}}{p_\mu}^2dp_\mu \int_{-1}^1d(cos\theta)\frac{1}{E_{\nu_\mu} E_\mu} L_{\mu\nu} J^{\mu\nu} Im{U_N(q_0, {\bf q})}.
\end{eqnarray}
where $L_{\mu\nu}=\sum L_\mu {L_\nu}^\dagger$ and ${J^{\mu\nu}}={\bar\sum}\sum J^\mu {J^\nu}^\dagger$. 

The leptonic current $L_{\mu}$ and the hadronic current $J^\mu$ are given by
\begin{equation}\label{lep_curr}
L_{\mu}=\bar{u}(k^\prime)\gamma_\mu(1-\gamma_5)u(k)
\end{equation}
\begin{equation}\label{had_curr}
J^\mu=\bar{u}(p^\prime)[F_{1}(q^2)\gamma^\mu + F_{2}(q^2)i{\sigma^{\mu\nu}}{\frac{q_\nu}{2M}} + F_{A}(q^2)\gamma^\mu\gamma_5 + F_{P}(q^2)q^\mu\gamma_5]u(p).
\end{equation}
where $q(=k-k^\prime)$ is the four momentum transfer, M is the mass of the nucleon, $G_{F}(=1.16637\times 10^{-5} GeV^{-2})$ is the Fermi coupling constant and $\theta$ is the lepton angle. $U_N$ is the Lindhard function for the particle hole excitation~\cite{Singh1}. The form factors $F_1$, $F_2$, $F_A$ and $F_P$ are isovector electroweak form factors and for our numerical calculations we have used the parameterisation of Bradford et al.~\cite{bradford} with axial dipole mass ${M}_{A}$=1.05GeV and vector dipole mass ${M}_{V}$=0.84GeV.
Inside the nucleus, the Q-value of the reaction and Coulomb distortion of outgoing lepton are taken into account by modifying the imaginary part of the Lindhard function $Im{U_N(q_0, {\bf q})}$ by $Im{U_N(q_0-V_c(r)-Q, {\bf q})}$. Furthermore, the renormalization of weak transition strength in the nuclear medium in a random phase approximation(RPA) is taken into account by considering the propagation of particle hole(ph) as well as delta-hole($\Delta h$) excitations. These considerations lead to modified terms involving the bilinear terms in the weak coupling constant in the hadronic tensor $J^{\mu\nu}_{RPA}$ for which expressions are given in Ref.~\cite{Athar2}. 
\subsection{INELASTIC RESONANCE PRODUCTION OF PIONS}
The basic reaction for the inelastic one pion production in nuclei, for a neutrino interacting with a nucleon inside a nuclear target is given by
\begin{equation}\label{inelastic_reaction}
\nu_{\mu}(k) + N(p) \rightarrow \mu^{-}(k^{\prime}) + N^{\prime}(p^{\prime}) + \pi^{+}(k_{\pi})~~~N, N^{\prime} = p/n 
\end{equation}
The cross sections for pion production is calculated using the $\Delta$ dominance model. In this model, the weak hadronic currents interacting with the nucleons in the nuclear medium excite a $\Delta$ resonance which decays into pions and nucleons. The pions interact with the nucleus inside the nuclear medium before coming out. The final state interaction of pions leading to elastic, charge exchange scattering and the absorption of pions lead to reduction of pion yield. The nuclear medium effects on $\Delta$ properties lead to modification in its mass and width which have been discussed earlier by Oset et al.~\cite{Oset} to explain the pion and electron induced pion production processes from nuclei. 

In the local density approximation the expression for the total cross section for the charged current one pion production is given by
\begin{eqnarray}\label{sigma_inelastic}
\sigma &=& \frac{1}{(4\pi)^5}\int_{r_{min}}^{r_{max}}(\rho_{p}(r)+\frac{1}{9}\rho_{n}(r)) d\vec r\int_{Q^{2}_{min}}^{Q^{2}_{max}}dQ^{2} \int_{0}^{\infty}dk^{\prime} \int_{-1}^{+1}d(cos\theta_{\pi}) \int_{0}^{2\pi}d\phi_{\pi} \times \nonumber\\
&&\frac{\pi|\vec  k^{\prime}||\vec k_{\pi}|}{M E_{\nu}^2 E_{l}}\frac{1}{E_{p}^{\prime}+E_{\pi}\left(1-\frac{|\vec q|}{|\vec k_{\pi}|}cos(\theta_{\pi })\right)} \bar\sum \sum|\mathcal M_{fi}|^2
\end{eqnarray}
where the proton density $\rho_{p}(r)=\frac{Z}{A}\rho(r)$ and the neutron density $\rho_{n}(r)=\frac{A-Z}{A}\rho(r)$ with $\rho(r)$ as the nuclear density taken as 3-parameter Fermi Density taken from Ref.\cite{Vries}.
The transition matrix element $\mathcal M_{fi}$ is given by
\begin{equation}\label{matrix_element}
\mathcal M_{fi}=\sqrt{3}\frac{G_F a}{\sqrt{2}}\frac{f_{\pi N \Delta}}{m_{\pi}} \bar \Psi({\bf P}) k^{\sigma}_{\pi} {\mathcal P}_{\sigma \lambda} \mathcal O^{\lambda \alpha} L_{\alpha} u({\bf p})
\end{equation}
where $L^{\alpha}$ is the leptonic current defined by Eq.(\ref{lep_curr}), $a=cos\theta_c$ and $\mathcal O^{\beta \alpha}={\mathcal O}_V^{\beta \alpha}+{\mathcal O}_A^{\beta \alpha}$ for the charged current induced $\pi^{\pm}$ production process while for the neutral current induced $\pi^{0}$ production process a=1 and $\mathcal O^{\beta \alpha}=(1-2 sin^{2}\theta_{W}){\mathcal O}_V^{\beta \alpha}+{\mathcal O}_A^{\beta \alpha}$. ${\mathcal O}_V^{\beta \alpha}$ and ${\mathcal O}_A^{\beta \alpha}$ are the vector and axial vector N-$\Delta$ transition operators given by 
\begin{equation}\label{vec_tra_current}
{\mathcal O}_V^{\beta \alpha}=\left(\frac{C_{3}^V(q^2)}{M}(g^{\alpha \beta} \not q-q^{\beta}\gamma^{\alpha})+\frac{C_{4}^V(q^2)}{M^2}(g^{\alpha \beta} q \cdot P-q^{\beta}P^{\alpha})+\frac{C_{5}^V(q^2)}{M^2}(g^{\alpha \beta}q \cdot p-q^{\beta}p^{\alpha}) \right) \gamma_{5}
\end{equation}
and
\begin{equation}\label{ax_tra_current}
{\mathcal O}_A^{\beta \alpha}=\frac{C_{4}^{A}(q^2)}{M^2}(g^{\alpha \beta}{\not q} -q^{\beta}\gamma^{\alpha})+C_{5}^{A}(q^2)g^{\alpha \beta}+\frac{C_{6}^{A}(q^2)}{M^2}q^{\beta}q^{\alpha}
\end{equation}
where ~$C_{i}^{V}(q^2)$ and  $C_{i}^{A}(q^2)$ are the vector and axial vector transition form factors and for our numerical calculations these have been taken from the work of Lalakulich et al.~\cite{Lalakulich}. $\theta_{W}$ is the weak mixing angle. ${\mathcal P}^{\sigma \lambda}$ is the $\Delta$ propagator in momentum space given by 
\begin{equation}	\label{width}
{\mathcal P}^{\sigma \lambda}=\frac{{\it P}^{\sigma \lambda}}{P^2-M_\Delta^2+iM_\Delta\Gamma}
\end{equation}
where ${\it P}^{\sigma \lambda}$ is the spin-3/2 projection operator given by
\begin{equation}\label{propagator}
{\it P}^{\sigma \lambda} = \sum_{spins} \psi^{\sigma} \bar \psi^{\lambda} = (\not P+M_{\Delta})\left(g^{\sigma \lambda}-\frac{2}{3} \frac{P^{\sigma}P^{\lambda}}{M_{\Delta}^2}+\frac{1}{3}\frac{P^{\sigma} \gamma^{\lambda}-P^{\sigma} \gamma^{\lambda}}{M_{\Delta}}-\frac{1}{3}\gamma^{\sigma}\gamma^{\lambda}\right)
\end{equation}
and the delta decay width $\Gamma$ is taken to be an energy dependent P-wave decay width taken as~\cite{Oset}:
\begin{equation}\label{Width}
\Gamma(W)=\frac{1}{6 \pi}\left(\frac{f_{\pi N \Delta}}{m_{\pi}}\right)^2 \frac{M}{W}|{\bf q}_{cm}|^3
\end{equation}
$|q_{cm}|$ is the pion momentum in the rest frame of the resonance and W is the center of mass energy.

Inside the nuclear medium the mass and width of delta are modified which in the present calculation are taken into account by using a modified mass $M_{\Delta} \rightarrow M_{\Delta}+ {Re}\Sigma_{\Delta}$ and modified width $\Gamma_{\Delta} \rightarrow \tilde\Gamma_{\Delta} - 2{Im}\Sigma_{\Delta}$ from the model developed by Oset et al.~\cite{Oset}, where $\tilde\Gamma_{\Delta}$ is reduced width of $\Delta$ due to Pauli blocking of nucleons in the $\Delta \rightarrow {N} \pi$ decay and $\Sigma_{\Delta}$ is the self energy of $\Delta$ calculated in nuclear many body theory using local density approximation. The expressions of ${Re}\Sigma_{\Delta}$ and ${Im}\Sigma_{\Delta}$ are taken from the Ref.~\cite{Oset}. The pions produced in this process are scattered and absorbed in the nuclear medium. This is treated in a Monte Carlo simulation which has been taken from the Ref.~\cite{Vicente}.
\subsection{DEEP INELASTIC REACTION}
The basic process for a neutrino interacting with a nucleon inside the nucleus is 
\begin{equation}\label{deepinelastic_reaction}
\nu_{\mu}(k) + N(p) \rightarrow \mu^{-}(k^{\prime}) + X(p^{\prime}). 
\end{equation}
where X is the jet of partons.

The differential scattering cross section for the deep inelastic scattering of (anti)neutrinos from unpolarized nucleons in the limit of lepton mass $m_l \rightarrow 0$, is described in terms of three structure functions, $F^\nu_1$($x$,$Q^2$), $F^\nu_2$($x$,$Q^2$) and $F^\nu_3$($x$,$Q^2$), where $x=\frac{Q^2}{2M\nu}=-\frac{q^2}{2M\nu}$ is the Bjorken variable, $\nu$ and $q$ being the energy and momentum transfer of leptons. In the asymptotic region of Bjorken scaling i.e. $Q^2 \rightarrow \infty$, $\nu \rightarrow \infty$, $x$ finite, all the structure functions depend only on the Bjorken variable $x$. In this scaling limit, $F^\nu_1(x)$ and $F^\nu_2(x)$ are related by the Callan-Gross relation~\cite{Callan} leading to only two independent structure functions $F^\nu_2$($x$) and $F^\nu_3$($x$) which are determined from the experimental data on deep inelastic scattering of (anti)neutrinos in the asymptotic region.

We have studied nuclear medium effects on the nucleon structure function $F^A_3$($x$,$Q^2$) in iron using spectral function to describe the momentum distribution of nucleons in the nucleus. The spectral function has been calculated using the Lehmann's representation for the relativistic nucleon propagator and nuclear many body theory is used to calculate it for an interacting Fermi sea in nuclear matter. A local density approximation is then applied to translate these results to finite nuclei~\cite{Marco}. Here we consider the modifications of nucleonic contributions to $F^A_3$($x$,$Q^2$) arising due to binding energy, off mass shell and Fermi motion of the nucleon in the nuclear medium which dominate in the region of $x\ge$0.3. In this model, in the region of 0.3$>x>$0.1, corresponding to the anti-shadowing region, the nuclear medium modification effects on $F^A_3$($x$,$Q^2$) are expected to be small due to vanishing of the pion contribution and we have not considered the shadowing region of 0.0$<x<$0.1. Therefore, our results should be able to describe the dominant contribution of nuclear medium effects to $F^A_3$($x$,$Q^2$) in the range of 0.1$<x<$1. 

The average structure function $F_{3}^{N}(x)$ on isoscalar nucleon target defined as
\begin{eqnarray*}\nonumber
F_{3}^{N}(x)=\frac{1}{2}\left(F^{\nu N}_{3} + F^{\bar\nu N}_{3}\right)
\end{eqnarray*}
is given by $
F_{3}^{N}(x)=[ u_v(x) + d_v(x) + s(x)- {\bar s}(x) + c(x) - {\bar c}(x)],
$
where $u_v(x)$ and $d_v(x)$ are the valence quark parton distributions. For an isoscalar target and a symmetric sea, $F_{3}^{N}(x)$ structure function is given in terms of valence quarks $u_v$ and $d_v$ which satisfy the Gross-Llewellyn Smith sum rule~\cite{GLS}:
\begin{equation}	\label{gls}
\int_0^1 F^N_3(x)dx=3.
\end{equation}
In the local density approximation the reaction given by Eq.(\ref{deepinelastic_reaction}) takes place at a point ${\bf r}$, lying inside the nucleus in a volume element $d^3r$ with local density $\rho_{p}({\bf r})$ and $\rho_{n}({\bf r})$ corresponding to the proton and neutron. The expression for $F^A_3 (x, Q^2)$ in the nuclear medium is given by~\cite{Athar5}:
\begin{eqnarray} 	\label{finalF3}
F^A_3 (x, Q^2)&=& 4 \int d^3 r \; \int \frac{d^3 p}{(2 \pi)^3} 
\int^{\mu}_{- \infty} \; d p^0 S_h (p^0, {\bf p}, \rho(r))
F(p, Q^2) F^N_3(x_N, Q^2),
\end{eqnarray}
where $x_N$ is the Bjorken variable expressed in terms of the nucleon variables, $(p^0, {\bf p})$, in
the nucleus i.e. $x_N=\frac{Q^2}{2p.q}$,

\[F(p, Q^2)=\frac{M}{E({\bf p})}\left(\frac{p_0 \gamma-p_z}{(p_0-p_z\gamma) \gamma}\right);~\gamma=\left(1+\frac{4M^2x^2}{Q^2}\right)^{1/2}~~~and\]

$S_h (\omega, {\bf p})$ is the hole spectral function, the expression for which is taken from Ref.~\cite{Fernandez}.
\section{RESULTS AND DISCUSSION}
\subsection{QUASIELASTIC REACTION}
In Fig.1, we present the ratio R of the charged current quasielastic lepton production cross section to the cross section on free nucleon defined as $R=\frac{1}{N}\frac{\sigma(^{12}C)}{\sigma(free)}$ as a function of neutrino(Fig.1a) and antineutrino(Fig.1b) energies when $\sigma(^{12}C)$ is calculated using Eq.(\ref{sigma_quasi}). We find that with the incorporation of nuclear medium effects without the RPA correlations the reduction in the cross section is around $45\%$ at $E_{\nu}$=0.2GeV, $16\%$ at $E_{\nu}$=0.4GeV, $10\%$ at $E_{\nu} \approx$ 1.0GeV and $7\%$ at $E_{\nu}$=2GeV from the cross sections calculated for the free case. However, when we also incorporate the RPA effects, the total reduction in the cross section is around $40\%$ at $E_{\nu}$=0.4GeV, $20\%$ at $E_{\nu}$=1.0GeV, $18\%$ at $E_{\nu}$=2GeV from the cross sections calculated for the free case. In the case of antineutrinos these reductions are larger as shown in Fig.1(b). We have compared our results with the results obtained in the Fermi gas model which has been used in the NUANCE Monte Carlo generator~\cite{nuance} by the MiniBooNE collaboration~\cite{boone1}. We find that the present results in the local Fermi gas model are similar to the results used in the NUANCE generator, but when RPA effects are included the cross sections are reduced. 
\begin{figure}
\includegraphics{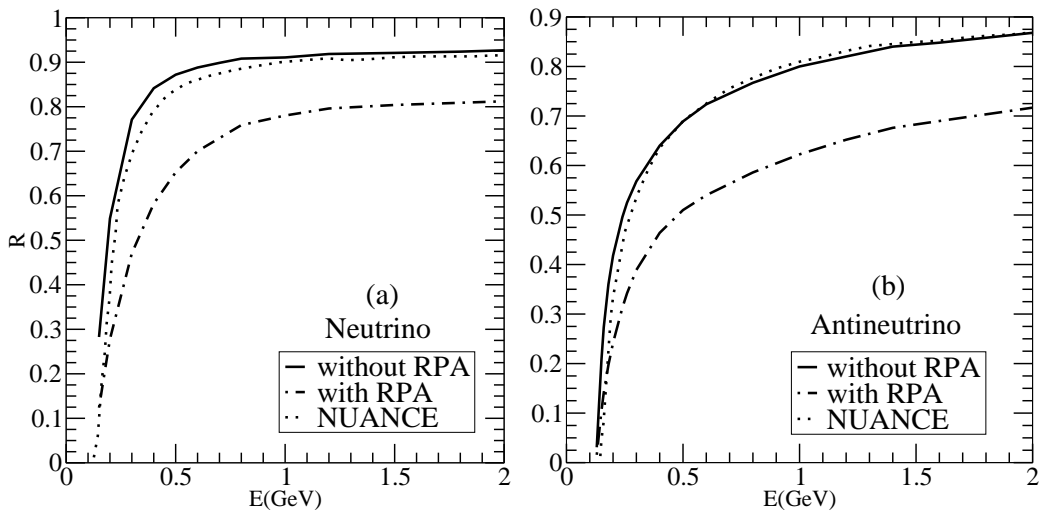}
\caption{$ R =\frac{1}{N}\frac{\sigma(^{12}C)}{\sigma(free)} $ vs Neutrino (Antineutrino) Energy.}
\end{figure}

\begin{figure}
\includegraphics{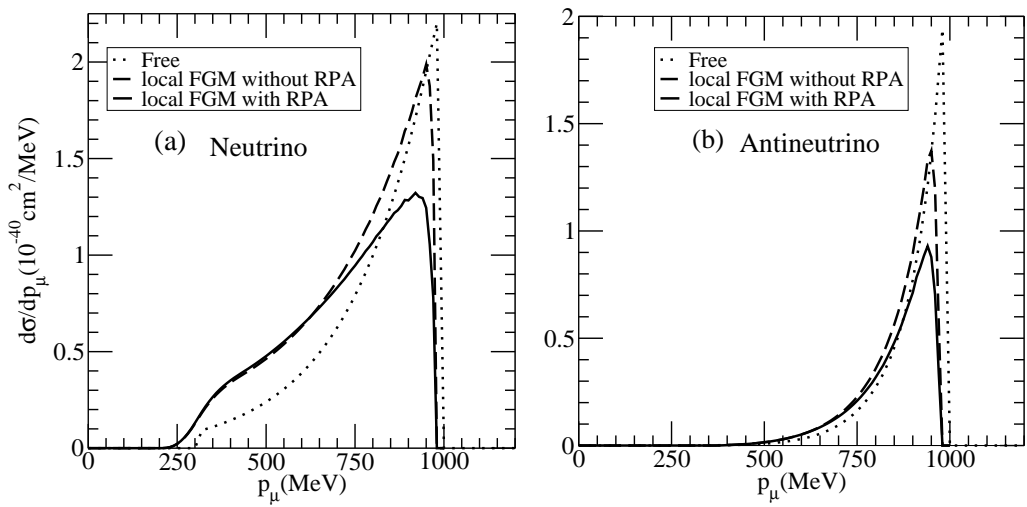}
\caption{$\frac{d\sigma}{dp_\mu}$ vs $p_\mu$ for the $\nu_\mu$(${\bar\nu}_\mu$) induced reactions on $^{12}C$ target at $E_{\nu}=1GeV$.}
\end{figure}

In Fig.2, we have shown the results for the lepton momentum distribution $\frac{d\sigma}{dp_{\mu}}$ for the $\nu_{\mu}$ and $\bar \nu_{\mu}$ induced charged current quasielastic processes. We find that when nuclear medium effects are taken into account there is a reduction as well as shift in the peak region towards the lower value of lepton momentum. This reduction in $\frac{d\sigma}{dp_{\mu}}$ when calculated in the local Fermi gas model without the RPA correlation effects as compared to the cross section calculated without the nuclear medium effects is around $10\%$ in the peak region of lepton momentum, which further reduces by around $30\%$ when RPA effects are also taken into account. In the case of antineutrino the reduction in $\frac{d\sigma}{dp_{\mu}}$ in the local Fermi gas model is around $30\%$ which further reduces by $30\%$ when RPA effects are also taken into account.
\subsection{INELASTIC REACTION}
In Fig.3, we present the results for $Q^2$-distribution $\frac{d\sigma}{dQ^2}$ and momentum distribution $\frac{d\sigma}{dp_{\pi}}$ for the charged current $\nu_\mu$(${\bar\nu}_\mu$) induced one $\pi^{+}$($\pi^{-}$) production cross section. These results are presented for the differential scattering cross section calculated with and without the nuclear medium effects and with nuclear medium effects including the pion absorption effects. For the $Q^2$- distribution shown in Fig.3a, we find that the reduction in the cross section as compared to the cross section calculated without the nuclear medium effects is around $35\%$ in the peak region. When pion absorpion effects are also taken into account there is a further reduction of around $15\%$. The results for the antineutrino induced one $\pi^{-}$ production cross section are qualititatively similar in nature but quantitatively we find that the peak shifts towards a slightly lower value of $Q^2$. In Fig.3b, the results for the pion momentum distribution have been shown. We find that in the peak region the reduction in the cross section is around $40\%$ when nuclear medium effects are taken into account, which further reduces by about $15\%$ when pion absorption effects are also taken into account. 
\begin{figure}
\includegraphics{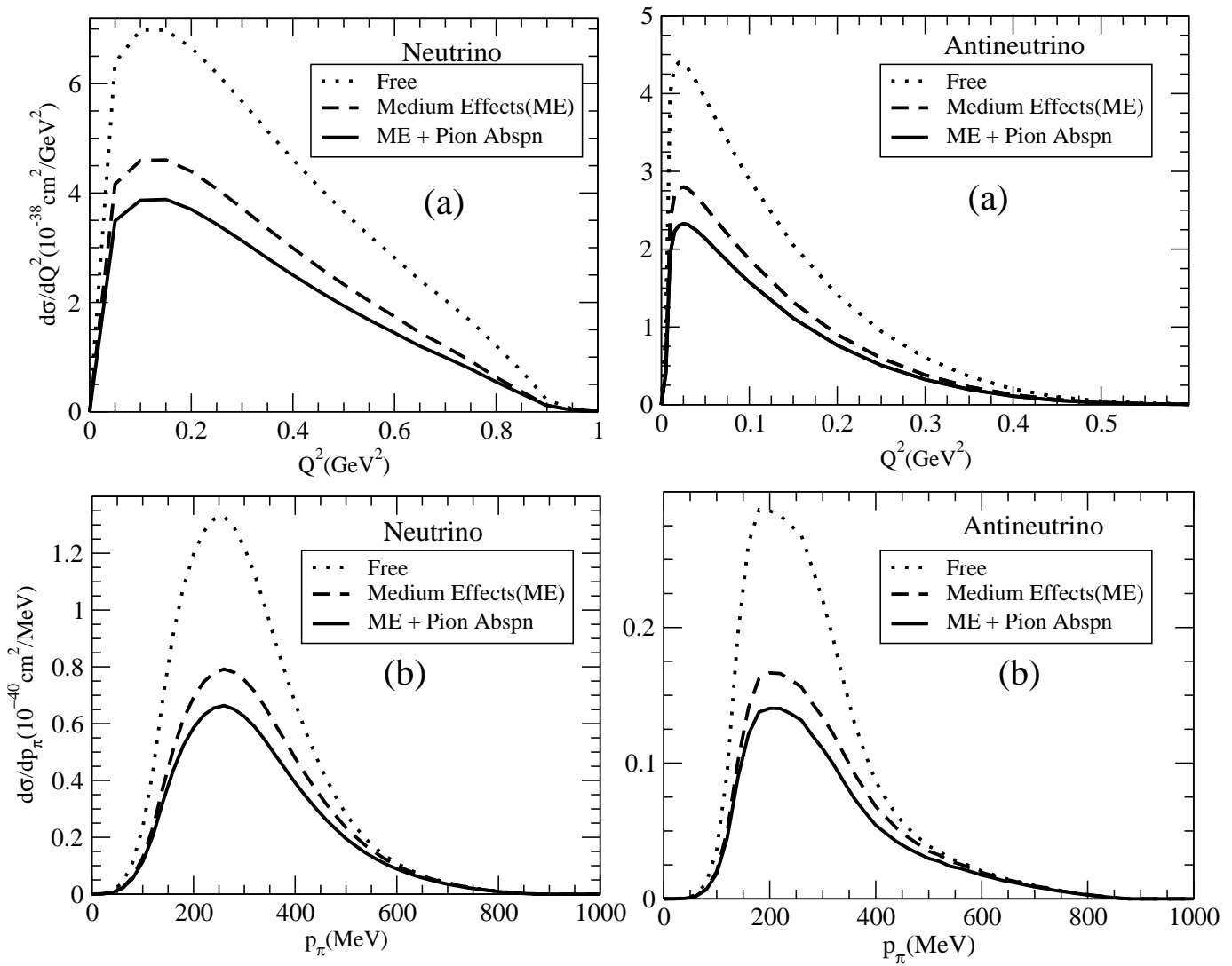}
\caption{$\frac{d\sigma}{dQ^2}$ and $\frac{d\sigma}{dp_\pi}$ for the $\nu_\mu$(${\bar\nu}_\mu$) induced charged current one $\pi^+(\pi^-)$ process on $^{12}C$ target at $E_{\nu}=1GeV$.}
\end{figure}
\begin{figure}
\includegraphics{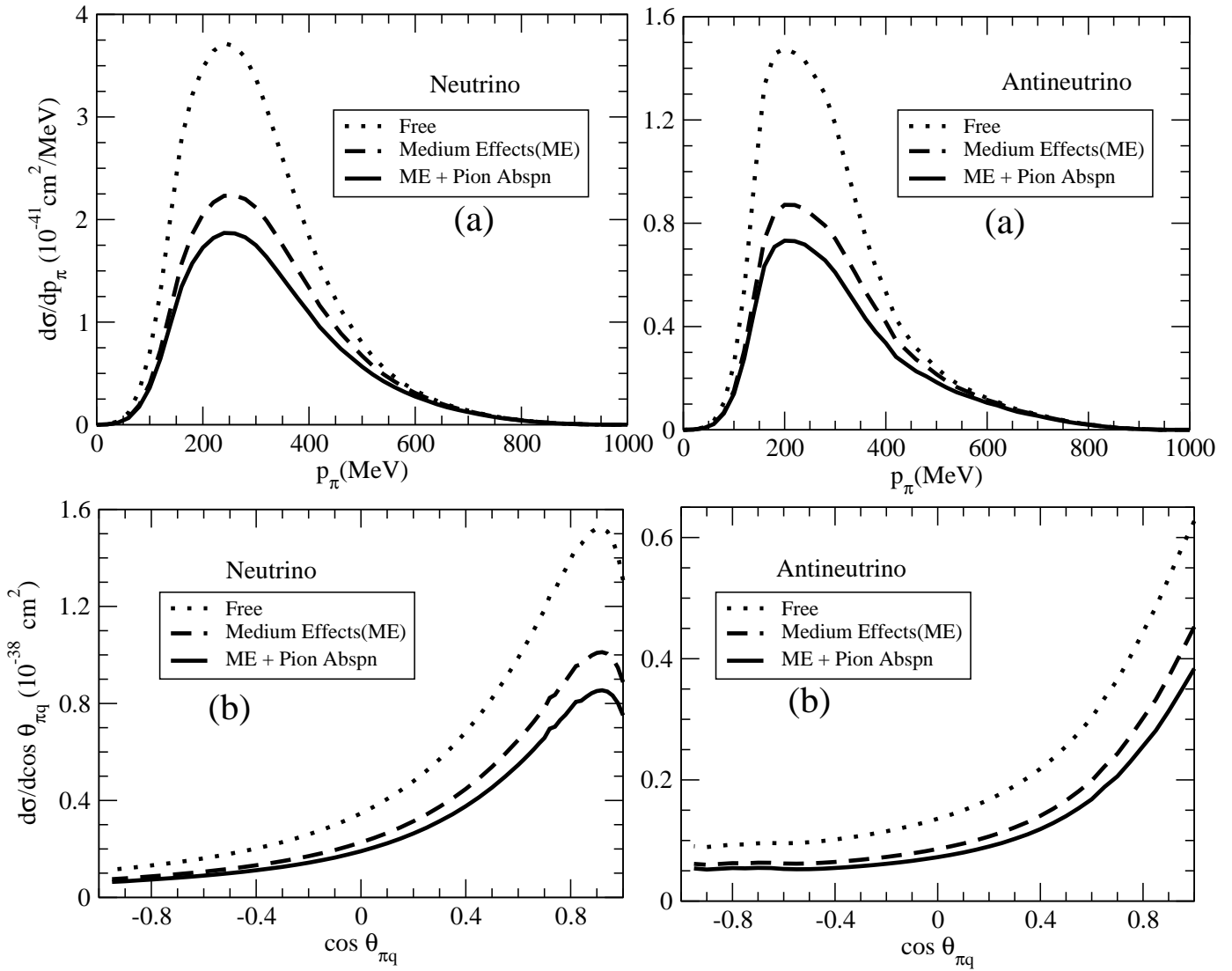}
\caption{$\frac{d\sigma}{dp_\pi}$ and $\frac{d\sigma}{dcos_{\theta q}}$ for the neutal current neutrino(antineutrino) induced $\pi^0$ production on $^{12}C$ target at $E_{\nu}=1GeV$.}
\end{figure}
\begin{figure}
\includegraphics{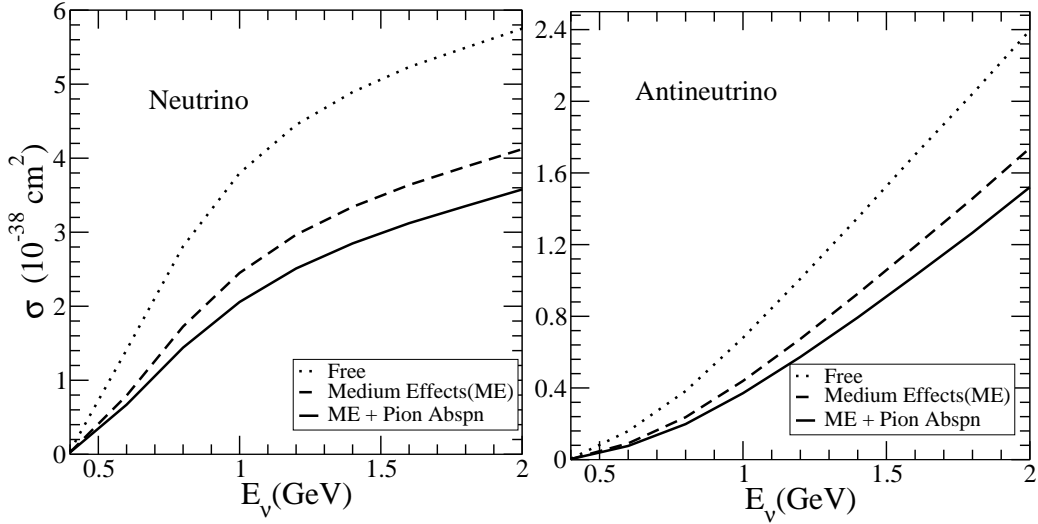}
\caption{$\sigma$ for $\nu_\mu$ (${\bar\nu}_\mu$) induced charged current one $\pi^+(\pi^-)$ production on $^{12}C$ target.}
\end{figure}
\begin{figure}
\includegraphics{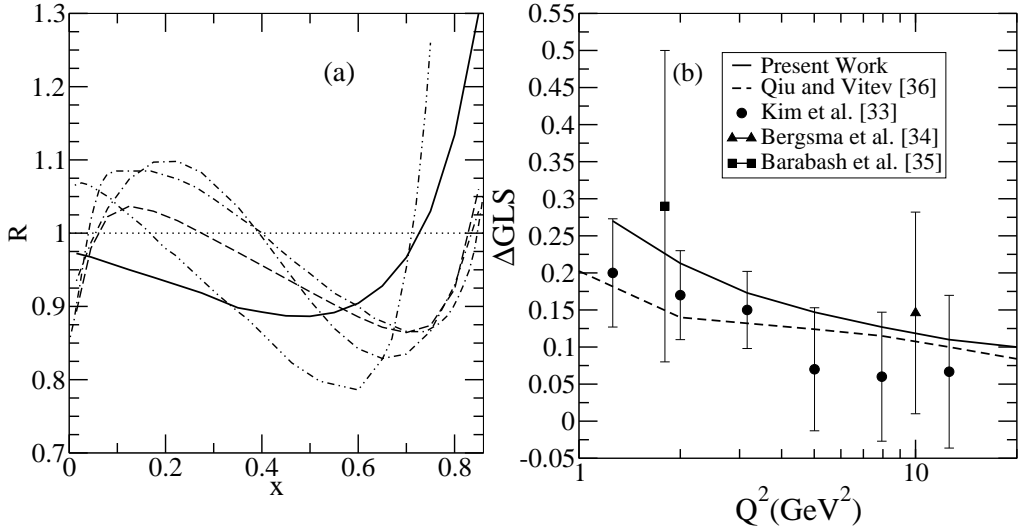}
\caption{(a)Results for the ratio $R=\frac{F_3^A(x)}{A F_3^N(x)}$ at $Q^2 = 5 GeV^2$ by different authors. Solid line: Result of the present calculation using MRST2004 NNLO parton distribution function; short dashed line: NuTeV Collaboration~\cite{Tzanov}, dashed-double dotted line is the result of Kulagin~\cite{Kulagin}, double dashed-dotted line: Hirai et al.~\cite{Hirai} and the results of Kulagin and Petti~\cite{Petti1} are shown by dashed-dotted line.(b)$\Delta GLS=\frac{1}{3}(3-\int_{0}^{1}F_{3}^{A}(x,Q^2)dx)$ vs $Q^2$.}
\end{figure}
In Fig.4, we present the results for the neutral current $\nu$($\bar \nu$) induced one $\pi^0$ production cross section. These results are presented for the pion momentum $\frac{d\sigma}{dp_{\pi}}$ and angular distributions $\frac{d\sigma}{dcos\theta_{\pi}}$, with and without the nuclear medium effects and with nuclear medium and pion absorption effects. For the pion momentum distribution shown in Fig.4a, we find that in the peak region the reduction in the cross section is around $40\%$ when nuclear medium effects are taken into account, which further reduces by about $15\%$ when pion absorption effects are also taken into account. The results with antineutrinos are similar in nature, except that in the case of $\bar \nu$, the angular distribution are more forward peaked than in the case of $\nu$. 

In Fig.5, we present the results for the total scattering cross section $\sigma$ for charged current $\nu_\mu$(${\bar\nu}_\mu$) induced one $\pi^{+}$($\pi^{-}$) production cross section. These results have been presented for the cross sections calculated without(with) the nuclear medium effects and also when pion absorption effect is included along with the nuclear medium effects. We find that with the inclusion of nuclear medium effects the reduction in the cross section from the cross section calculated without the nuclear medium effects for neutrino energies between 1-2 GeV is 30-35$\%$ which further reduces by 15$\%$ when pion absorption effects are also taken into account. The results with antineutrinos are similar in nature.
\subsection{DEEP INELASTIC REACTION}
In Fig.6(a), we compare our results for R($x$,$Q^2$) at $Q^2=5GeV^2$, where $R=\frac{F^A_{3} (x,Q^2)}{A F^N_{3} (x,Q^2)}$, with the results of Tzanov et al.~\cite{Tzanov}, Kulagin and Petti~\cite{Petti}, Kulagin~\cite{Kulagin} and Hirai et al.~\cite{Hirai}. While the work of Kulagin~\cite{Kulagin} and Kulagin and Petti~\cite{Petti,Petti1} use a nuclear model to calculate the nuclear effects which shows a $Q^2$ dependence, the work of Tzanov et al.~\cite{Tzanov} and Hirai et al.~\cite{Hirai} are phenomenological analyses, which assume the nuclear
effects to be independent of $Q^2$. We find a
suppression in $F^A_3(x, Q^2)$ for $x<$0.7 and an enhancement thereafter,
 which are respectively smaller than the results of Kulagin~\cite{Kulagin}, but are larger than the recent results of Kulagin
and Petti~\cite{Petti1}. It should be noted that
these latter results~\cite{Petti1} give  suppression in the region of 0.4$<x<$0.8 and enhancement for $x>$0.8, which
are smaller than the present results and the results obtained earlier in Ref.~\cite{Kulagin}. When compared with the results of Tzanov et al.~\cite{Tzanov} and Hirai et al.~\cite{Hirai}, we find a smaller suppression in the region 0.5$<x<$0.7. In the region 0.7$<x<$0.8, we find an enhancement while they obtain a suppression. In Fig.6(b), we show the $Q^2$ dependence of the nuclear effects of the GLS integral, where we plot $\Delta$GLS=$\frac{1}{3}(3-\int_0^1F^A_3(x, Q^2)dx)$ as a function of $Q^2$. The experimental results from CCFR collaborations~\cite{Kim}, CHARM collaborations~\cite{Bergsma} and IHEP-JINR collaborations~\cite{Barabash} are also shown. The $Q^2$ behaviour of $\Delta$GLS has been found to be in reasonable agreement with the present available experimental results. In this figure, we have also shown the theoretical results obtained by Qiu and Vitev~\cite{Qiu}. Our results are in agreement with the results of Qiu and Vitev~\cite{Qiu} for $Q^2>5GeV^2$ where theoretically the suppression is found to be larger than the experimental results. For $Q^2<5GeV^2$, we find a larger suppression compared to the central value of the experimental result and both theoretical values are within the experimental errors. 
\section{CONCLUSIONS}
We will like to conclude that :

(i) In the case of charged current quasielastic lepton production, the role of nuclear medium effects like Pauli blocking, Fermi motion is to reduce the cross section. When nuclear correlations in the nuclear medium are taken into account in a Random Phase Aprroximation (RPA) there is further reduction in the cross section. The total reduction in the cross section is about $20\%$ in case of $\nu$ and slightly larger for $\bar \nu$ around $E_\nu$ = 1GeV. 

(ii) In the case of charged current one pion production, the nuclear medium and pion absorption effects lead to a reduction in the cross section about 45$\%$ at $E_{\nu}$=1 GeV and give appreciable distortion in the energy and angular distribution of pions and leptons.

(iii) In the case of deep inelastic scattering, the nuclear effects decrease the value of the structure function $F^A_{3} (x, Q^2)$ in the iron nucleus for $x \le x_{min}$=0.7 and increase it at higher $x>x_{min}$. In general nuclear medium effects decrease the value of GLS integral for all $Q^2$.
\begin{center}
{\bf ACKNOWLEDGMENTS}
\end{center}
 The work presented here is mainly supported by Government of India through the grant DST-SP/S2K-07/2000 and the academic exchange program between Aligarh Muslim University, Aligarh and University of Valencia, Spain. S. C. would like to thank the Jawaharlal Nehru Memorial Fund for the Doctoral Fellowship.

\end{document}